\documentclass[prl,twocolumn,showkeys,showpacs,preprintnumbers,amssymb,amsmath,superscriptaddress]{revtex4}

\usepackage{graphicx}
\usepackage{dcolumn}
\usepackage{bm}
\usepackage{amssymb,amsmath}

\usepackage{color}
\usepackage{ulem} 

\begin{document}


\title{Space-charge modulation in vacuum microdiodes at THz frequencies}

\author{A. Pedersen}
\author{A. Manolescu}
\author{\'{A}. Valfells}
\affiliation{School of Science and Engineering, Reykjavik University, 103 Reykjav\'{\i}k, Iceland}

\date{\today}

\begin{abstract}

We investigate the dynamics of a space-charge limited, photoinjected, electron beam in a microscopic vacuum diode.  Due to the small nature of the system it is possible to conduct high-resolution simulations where the number of simulated particles is equal to the number of electrons within the system.  In a series of simulations of molecular dynamics type, where electrons are treated as point-charges, we address and analyze space-charge effects in a $\mu$m-scale vacuum diode.  We have been able to reproduce breakup of a single pulse injected with a current density beyond the Child-Langmuir limit, and we find that continuous injection of current into the diode gap results in a well defined train of electron bunches corresponding to THz frequency.  A simple analytical explanation of this behavior is given.

\end{abstract}

\pacs{
52.25.Tx 
52.59.Rz 
29.20.Ba 
29.27.Bd 
} 


\maketitle


Vacuum microelectronic devices have shown some promise as sources of radiation with frequency ranging from GHz to THz.  Examples of this include microtriodes ~\cite{bower1,bower2}, nanoklystrons ~\cite{siegel} and traveling wave tubes ~\cite{bhattacharjee}.  Most such devices rely upon field emission as a source of electrons, although photoemission could also have similar advantages in terms of cold-cathode behavior.  In either case, it is reasonable to revisit some of the basic physics that describe macroscopic vacuum electronic devices and to see how they might manifest differently on the microscale.  One example is the problem of space-charge limited flow in a diode.  The classic Child-Langmuir equation ~\cite{child,langmuir} describes this to an extent.  It is derived assuming continuous flow of electrons from the cathode to the anode in a planar diode of infinite width.  The initial analysis has since been improved upon to include the effects of limited emission area ~\cite{luginsland, lau} finite pulse length ~\cite{valfells}, and quantum effects in small diode gaps ~\cite{ang}, to name a few.  These improvements have been concerned with finding the onset of virtual cathode formation, so as to determine the upper limit of current flow that can be smoothly transmitted through a vacuum diode.  In this study we look at the dynamics of the beam once the space-charge limit has been attained.  


\begin{figure}
\includegraphics[width=8.5cm]{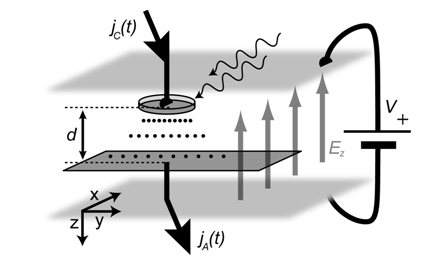}
\caption{Schematic model of the system simulated. The variable parameters are the applied voltage, $V$, and the gap width, $d$. The resulting currents at the cathode (emitter), ${j_C(t)}$, and at the anode (absorber), ${j_A(t)}$, are measured.} \label{fig:schematicModel}
\end{figure}

The model used consists of a planar cathode with a circular emitter area of radius of 250 nm seperated from a planar anode of infinite width by a vacuum gap of length $d$. Due to the small size of the system under study, it is possible to treat all electrons within the system individually.
An {\it external} static electric potential ${V}$ is applied across the gap an gives rise to an electric field, $E_{z}=V/d$, which accelerates the electrons from the cathode towards the anode. In addition to the external field an electron, $i$, also experiences the {\it internal} Coulomb field due to the space-charge created by the other electrons:
\begin{align}
\label{equ:Coulomb} {\bf E}_{sc}=-\frac{1}{4\pi\epsilon_{0}}\sum_{j\neq i}^{N}{\frac{e}{\left|{\bf r}_{ij}\right|^2}\hat{{\bf r}}_{ij}}
\end{align}
where $\epsilon_{0}$ is the permittivity of free space, $e$ is the electron charge, and the sum runs over all other electrons present in the gap. $r_{ij}$ is the relative distance between electron $i$ and $j$. Yielding the total field ${{\bf E}_{tot}={\bf E}_{sc}+[0,0,E_{z}}]$

Dynamic behavior is obtained by solving the classical (Newton) equations of motion using a numeric Velocity Verlet integrator with a discrete time step of 1 fs. During a simulation electrons continuously enter and leave the simulation box, such that the number electrons in the device is not conserved. Relativistic effects and radiation caused by acceleration of electrons are safely neglected as all occuring velocities are much smaller than the speed of light. The model is shown schematically in Fig.~\ref{fig:schematicModel}. Note that the ${z}$-coordinate is defined by the normal to the surface of the cathode.

The simulations conducted are divided into two groups according to the time profile of the laser pulse exciting photoelectrons.
(I) Pulse scenario, in which an electron pulse of duration less than the transit time is created by varying the laser intensity for a fixed applied potential, or varying the the external potential for a fixed laser intensity.
(II) Beam scenario, in which a continuous laser beam is emitted and either the gap length, or the external potential is varied. Reported values are from representative simulations for the specific parameter-set. However, all simulations have been repeated 10 times and only small variations are observed.

The emission model used in the simulations assumes that photoelectrons are available in proportion to the photon intensity.  Whether these photoelectrons are emitted is entirely dependent on whether the local electric field at their place of origin accelerates them away from the cathode or not.  For an electron to enter the system at the cathode it {\it must} experience a force towards the anode. With the definition of the $z$-direction as shown in Fig.~\ref{fig:schematicModel} the emission criterion becomes ${F_{tot,z} > 0}$. During the simulation an electron attempts to enter the system at a random point on the cathode where the condition  ${F_{tot,z} > 0}$ is tested by placing the electron 1 nm behind the surface. If true, the particle is promoted to the surface and then the test is repeated.  If the result is again true the electron is finally emitted, with zero initial velocity. 
Unsuccessful emission attempts are registered and a series of 100 unsuccessful attempts will stop the emission process. The inspection of ${F_{tot,z}}$ behind the surface ensures that all electrons already placed on the cathode are observed. We note that by varying the depth of the initial test the surface-density is affected. This depth can be viewed as being related to the skin-depth of the cathode material.


\begin{figure}
\includegraphics[width=8.5cm]{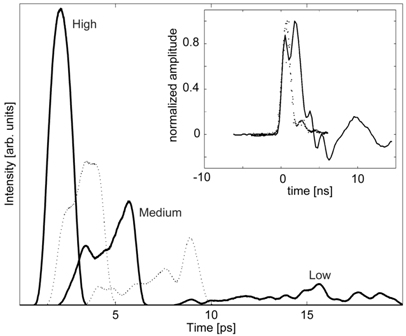}
\caption{
Shape of the electron pulse when the externally applied voltage is varied. A fixed Gaussian photon profile is used. Only the highest voltage gives a 100\% yield.
Inset is earlier published experimental results, reproduced from Ref.~\cite{valfells}.
The experimentally obtained double peaked shape was assigned to space-charge effects, this structure is recovered for the Medium voltage.} \label{fig:varyVoltage}
\end{figure}

In earlier experimental investigations by Valfells {\it et al.}~\cite{valfells} space-charge effects have been examined for photoelectrons extracted from a Pierce gun using a laser pulse with a Gaussian-like intensity profile in time. In our simulations a radial electric field, $E_{r}=2Vr/d^2$, represents the confinement of the Pierce gun, where $r$ is the radial distance to the center of cathode and $d$ equals the gap length that is kept constant at 2~$\mu$m.
The intensity of a monochromatic photon pulse can be associated to the time interval between consecutive emission attempts, and a Gaussian time-interval profile can thus be applied to reproduce the Gaussian intensity profile. The pulse duration is defined as the time interval between first and last emission attempt and it is kept constant at 1~ps, whereas the total number of photons in a pulse (intensity) is a variable parameter. For the short-pulse limit under investigation, the transit time of the electrons between the cathode and the anode exceeds the pulse duration, $t_{pulse}<<t_{transit}.$ Two series of pulse simulations are carried out where either the intensity of the photon pulse or the applied external potential is varied.
When the intensity is varied, three regimes are simulated while the driving and confinement potential are kept constant at 2~V. For these parameters it is found that a virtual cathode is formed if the photon pulse exceeds a total of 100~photons.  The nomenclature we use, is to call a pulse with 100~photons a low intensity pulse. All higher intensities will result in yields of electrons less than 100\%. For the medium intensity the number of photons in the pulse is increased by one order of magnitude and for the high intensity by two orders of magnitude. The main effect of the space-charge for a photon pulse of low intensity is a stretching of the electron pulse that is expected due to the B\"orsch effect. It should be noted that only for the low intensity do the peak value of the photon pulse and electron pulse coincide. For the medium and high photon pulse it is only the initial part that causes emission. The reason for this is that space-charge accumulation in the gap blocks subsequent electrons from being emitted, and the obtained electron pulse becomes a truncated Gaussian profile. Significant coupling in the $z$-direction and momentum transfer causes the electron pulse to spread out in both time and in velocity space when the medium and high intensity photon pulse are applied. A characteristic of the mentioned effects is that the electron pulse develops into a doubled peaked structure.

For the simulations where the applied voltage is varied, a fixed pulse size of 350~photons is used. The applied potential is reduced from 10.0~V, labeled high in Fig.~\ref{fig:varyVoltage}, by an order of magnitude, labeled medium, and finally by two orders of magnitude, labeled low. The confinement potential is scaled accordingly. In the regime labeled high the external field dominates, whereas it becomes comparable to the internal field in the medium case, and for the low case the internal field is dominant.  The high potential conserves the structure of the exciting photon pulse when the electron pulse reaches the anode, although the electron pulse becomes slightly stretched. Also, only by applying the high potential is a 100\% yield obtained.
As can be seen in Fig.~\ref{fig:varyVoltage} do space-charge effects become significant when the applied potential decreases to the medium regime. At this value the electron pulse transforms into a double peaked structure and the yield decreases to 57\%. As the applied potential decreases further the Coulomb interactions becomes dominant. A result of this is that the initial shape of the photon pulse gets completely washed out and it widens fourfold. The yield is reduced to 24\%.

In conclusion, for both series of simulations the medium profiles are double-peaked structures. This characteristic of the electron pulse is consistent with experimental observations by Valfells {\it et al.}~\cite{valfells} where a distinct feature of the electron pulse in the regime where space-charge effects are significant is indeed a double-peaked structure.


\begin{figure}
\includegraphics[width=8.5cm]{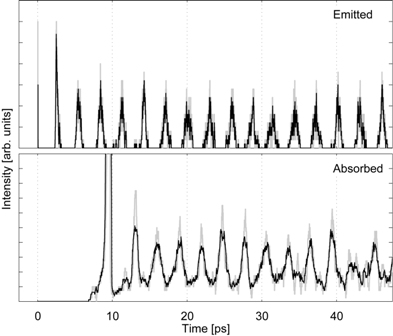}
\caption{s
Electron intensity as a function of time when a voltage of 1~V is applied over a 2~$\mu$m gap.
A continuous photon beam transforms into a pulsed signal of electrons with a frequency of 720~GHz.
An absorbed electron is attributed a Gaussian profile in time and for the black curve (grey curve) $\sigma=$ is 800~fs (400~fs).
Note that the frequency remains unaffected of profile-width and that the individual peaks are well separated.}
\label{fig:currents}
\end{figure}
\begin{figure}
\includegraphics[width=8.5cm]{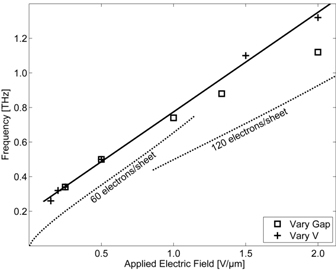}
\caption{
Frequency as a function of applied electric field.
The deviating behavior at small gap-lengths (large electric fields) arises when the transit time of a charge-sheet becomes comparable to the period of the density oscillation.
Solid line is a linear fit to the obtained data.
Dashed lines are from the proposed model given by Equ.~\ref{frq}.}
\label{fig:frequency}
\end{figure}

Next we look at the simulations where the cathode is subject to a continuous photon beam, $t_{transit}<<t_{beam}$. In these simulations the intensity of the photon beam is sufficient for the cathode to reach saturation. The condition for saturation is that the field created by the space-charge $E_{sc,z}$ exceeds the applied field everywhere on the cathode, $E_{z}<-E_{sc,z}$. As long as this situation holds the emission process is inhibited.

Simulations of a continuous photon beam exciting the cathode are conducted without any confining potential in the $xy$-plane. The parameters varied are either the gap length $d$ or the applied voltage $V$. In both cases a surprising effect is that the continuous photon beam generates a train of electron pulses with a well defined seperation in time corresponding to an oscillating current. These oscillations will be the main focus in the remainder of this paper. Representative results for the pulsed current at the cathode, ${j_C(t)}$, and at the anode, ${j_A(t)}$, are shown in Fig.~\ref{fig:currents}. All simulations with the applied electric field exceeding 0.1 V/$\mu$m show clear oscillations in the electron density as a function of time at the cathode. At the anode the internal repulsion between electrons has caused an averaging wash-out dependent on the transit time. It should be noted that the period of the oscillations is unaffected by this wash-out, but a significant decrease in the amplitude occurs.

In the first series of simulations the gap length is varied from 0.5~$\mu$m to 4.0~$\mu$m and a constant potential of 1~V is applied. Only the first density peak expands freely and the number of electrons in this initial peak is larger than for the following peaks. For the following peaks the number of electrons per pulse only shows small fluctuations. This latter observation indicates that the system quickly reaches a steady state. In Fig.~\ref{fig:frequency} one should note that the determined frequency bends off when the gap width becomes small, increasing the applied field $E$. The other series of simulations conducted varies the applied voltage from 0.2~V to 4.0~V while keeping the gap width fixed at 2~$\mu$m.  The correlation between the frequency of the density peaks and the applied field is shown in Fig.~\ref{fig:frequency}.


\begin{figure}
\includegraphics[width=8.5cm]{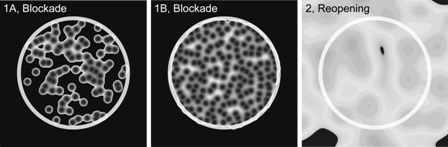}
\caption{
Map of cathode showing regions open for emission, black, whereas grey and white regions are blocked.
The circle marks the rim of the cathode.
Left (1A), a blockade is formed as electrons are added.
Middle (1B), the complete blockade has formed.
Right (2), a reopening of the emitter occurs when the electron-sheet exceeds a critical distance.}
\label{fig:electricField}
\end{figure}

To understand the density oscillations observed in Fig.~\ref{fig:currents} three phases of the cathode are of special importance: (1A) Creation of a charge-sheet.  (1B) Blockade, when a full charge-sheet has formed.  (2) Reopening.  For each of these phases we determined the total or effective $z$-component of the electric field at the cathode surface, which we define as
$E_{tot,z}=E_{z}+E_{sc,z}$.
The results are shown in Fig.~\ref{fig:electricField}, which is a map of the regions open for emission. From Fig.~\ref{fig:electricField}-1A and Fig.~\ref{fig:electricField}-1B it appears that the area blocked by a single electron limits to its vicinity within a discrete region on the surface of the cathode. Also to be noted is that the area blocked by an electron is unaffected by the presence of other electrons.

These two observations indicate that the total number of electrons in a charge sheet will be determined by the area fraction blocked by a single electron on the cathode surface. When the surface is fully covered, Fig.~\ref{fig:electricField}-1B, the emission of electrons stops and the cathode enters a blocked state. Regions distant from an electron are subject to an enhanced blockade when the newly emitted charge-sheet starts to move toward the anode. The enhancement arises because the blockade from an electron in a fixed point on the cathode depends both on the distance and on the angle to the surface.
As the distance increases further, the field from the charge sheet transforms from discrete to nearly continuous. Finally, when the charge sheet has covered a critical distance a reopening of the emitter occurs, Fig.~\ref{fig:electricField}-2. At this distance $E_{tot,z}$ changes sign and the formation of a new charge-sheet starts. Because the field distribution is now continuous the whole area reopens almost immediately. The sequence of these three states reveals that the period of the density oscillations is determined by the time it takes a charge-sheet to cover a critical distance.

There is no simple correlation between the number of electrons in a charge sheet, or neither the applied potential nor the gap length. 
The number is between [66, 123] when the potential is varied and [66, 218] when the gab length is varied, however, a clear trend in both cases is that the number of electrons increases as the $E_{z}$ increases. To understand this qualitatively it is reasonable to assume that the building time of a charge-sheet does not change significantly when $E_{z}$ is varied.  Also, the speed of the sheet depends mostly on the applied field. The extra electrons entering a charge-sheet as the applied potential increases is therefore a consequence of a larger distance traveled by the former blocking charge-sheet before the onset of the enhanced blockade.

With these assumptions we can derive a simplified analytical model for the density oscillations.  Suppose the cathode releases a thin sheet of charge which is then accelerated by the external field $E_{z}$. Assume that the charge sheet is a disk of radius $R$, like the cathode itself, situated at position $z$ in front of the cathode.  The field created at the center of the cathode is
$E_{sc,z}(z)=E_0\left(1-\frac{z}{\sqrt{z^2+R^2}}\right)$\,, where $E_0=\sigma/2\epsilon_0$ and $\sigma=-Ne/\pi R^2$ is the charge density of a sheet containing $N$ electrons.  The distance at which $E_{tot,z}=0$ in the center of the disk is then $z_0=R \left(1-E_{z}/E_0)\right / \sqrt{1-\left(1-E_{z}/E_0\right)^2}$\,.  Let us admit that the acceleration of the charge sheet due to the external field is constant, $a=eE_{z}/m$, $m$ being the electron mass. The time until the total field vanishes, and the cathode is ready to build the next sheet of charge, is $t_0=\sqrt{2mz_0/eE_{z}}$ and thus the frequency of the charge pulses is $f=1/t_0$: %
\begin{equation} \label{frq}
f=\sqrt{ \frac{eE_{z}}{2mR}} \frac {\left[ 1-\left(1-E_{z}/E_0\right)^2 \right]^{1/4}}
      {\left(1-E_{z}/E_0\right)^{1/2}} \,.
\end{equation}
If the external field is weak, $E_{z} \ll E_0$, then $f\approx\sqrt\frac{eE_{z}}{2mR} \left(\frac{2E_{z}}{E_0}\right)^{1/4} \sim E_{z}^{3/4}.$
Eq.\ ({\ref{frq}}) predicts that for a large external field, $E_z \to E_0$, the frequency diverges, or the oscillation period tends to zero.  In that case the Coulomb field is significantly weaker relatively to the external field and the cathode is never blocked.

The authors would like to thank for funding provided by the Reykjavik University Development Fund.


\begin{thebibliography}{12}

\bibitem{bower1} {C. Bower {\it et al.}
IEEE Trans. Electron Devices, {\bf 49}, 1478 (2002)}

\bibitem{bower2} {C. Bower {\it et al.}
Appl. Phys. Lett., {\bf 80}, 3820 (2002)}

\bibitem{siegel} {P.H. Siege {\it et al.}
Proc. of the 12th International Space Terahertz Technology Symposium, 81 (2001)}

\bibitem{bhattacharjee} {S. Bhattacharjee {\it et al.}
IEEE Trans. Plasma Science, {\bf 34}, 1002 (2004)}

\bibitem{child} {C.D. Child
Phys. Rev., {\bf 32}, 492 (1911)}

\bibitem{langmuir} {I. Langmuir
Phys. Rev., {\bf 21}, 419 (1923)}

\bibitem{luginsland} {J.W. Luginsland {\it et al.}
Phys. Rev. Lett., {\bf 77}, 4668 (1996)}

\bibitem{lau} {Y.Y. Lau
Phys. Rev. Lett., {\bf 87}, 278301 (2001)}

\bibitem{valfells} {\'{A}. Valfells {\it et al.}
Phys. Plasmas, {\bf 9}, 2377 (2002)}

\bibitem{ang} {L.K. Ang {\it et al.}
IEEE Trans. Plasma Science, {\bf 32}, 410 (2004)}

\end{thebibliography}

\end{document}